\documentclass[twocolumn,prd,showpacs,preprintnumbers,superscriptaddress,amsmath,amssymb,citeautoscript,longbibliography]{revtex4-1}
\pdfoutput=1
\usepackage{graphicx}
\usepackage{epstopdf}
\usepackage{dcolumn}
\usepackage{color}
\usepackage[dvipsnames]{xcolor}
\usepackage{bm}
\usepackage[hyperindex]{hyperref}
\hypersetup{
     colorlinks   = true,
     citecolor    = blue,
     urlcolor = blue,
     linkcolor = blue
}

\DeclareMathOperator{\sh}{sh}

\DeclareMathOperator{\cth}{cth}
\DeclareMathOperator{\sgn}{sgn}
\DeclareMathOperator{\tr}{tr}

\newcommand{\bpm}{\begin{pmatrix}}
\newcommand{\epm}{\end{pmatrix}}
\newcommand{\bs}{\boldsymbol}

\begin{document}

\title{RKKY interaction on the surface of three-dimensional Dirac semimetals}

\author{V. Kaladzhyan}
\email{vardan.kaladzhyan@phystech.edu}
\affiliation{Laboratoire de Physique des Solides, CNRS, Univ. Paris-Sud, Universit\'e Paris-Saclay, 91405 Orsay Cedex, France}
\affiliation{Department of Physics, KTH Royal Institute of Technology, Stockholm, SE-106 91 Sweden}

\author{A. A. Zyuzin}
\affiliation{Department of Applied Physics (LTL), Aalto University, P.O. Box 15100, FI-00076 AALTO, Finland}
\affiliation{Ioffe Physical-Technical Institute, 194021 St. Petersburg, Russia}

\author{P. Simon}
\affiliation{Laboratoire de Physique des Solides, CNRS, Univ. Paris-Sud, Universit\'e Paris-Saclay, 91405 Orsay Cedex, France}


\begin{abstract}
We study the RKKY interaction between two magnetic impurities located on the surface of a three-dimensional Dirac semimetal 
with two Dirac nodes in the band structure. 
By taking into account both bulk and surface contributions to the exchange interaction between the localized spins, we demonstrate that the surface contribution in general dominates the bulk one at distances larger than the inverse node separation due to a weaker power-law decay. We find a strong anisotropy of the surface term with respect to the spins being aligned along the node separation axis or perpendicular to it. In the many impurity dilute regime, this implies formation of quasi-one-dimensional magnetic stripes orthogonal to the node axis. We also discuss the effects of a surface spin-mixing term coupling electrons from spin-degenerate Fermi arcs. 
\end{abstract}


\maketitle

\section{Introduction}
Since the mid-twentieth century it has been understood that localized spins in metals can interact by means of the Ruderman-Kittel-Kasuya-Yosida (RKKY) mechanism \cite{Ruderman1954,Kasuya1956,Yosida1957}. This indirect exchange coupling is mediated by the conduction electrons and has been investigated in materials of different nature such as disordered metals \cite{Zyuzin1986}, superconductors \cite{Abrikosov1988, Aristov1997,Galitski2002}, topological insulators \cite{Liu2009,Biswas2010,Garate2010,Zhu2011,Abanin2011,Tokura2012,Zyuzin2014, Pankratov_RKKY},
 graphene \cite{Saremi2007,Hwang2008,Black-Schaffer2010,Sherafati2011,Kogan2011}, carbon nanotubes \cite{Braunecker2009prl,Klinovaja2013}, and semiconducting wires \cite{Braunecker2009}.

The RKKY interaction in Weyl and Dirac semimetals has been studied theoretically throughout recent years \cite{Sun2015,Chang2015,Hosseini2015,Sun2017}. To our knowledge, previous studies focused mainly on addressing interactions between localized spins immersed into the bulk of these exotic materials. However, to date the problem of magnetic impurities localized on the surface of 3D Weyl and Dirac materials has not been addressed.

There exists several strong incentives of studying the RKKY interaction on the surface of a Dirac material. 
The main motivation is that 3D Dirac materials have been found  and characterized experimentally
in Na$_3$Bi and Cd$_3$As$_2$ \cite{Liu2014,Borisenko2014,Neupane2014,Jeon2014,Xu2015}.
Second, surfaces are often imperfect and contain impurities, some being magnetic. The interactions between magnetic impurities may affect and sometimes dominate certain types of measurements, including the widespread scanning tunneling microscopy. Third, the surface states of such materials, the so-called Fermi arc states, contain anisotropically dispersed electrons. Furthermore, the node separation lines provide specific spatial anisotropy in these materials. Such peculiar features may give rise to a nontrivial  RKKY interaction with a strongly direction-dependent asymptotic behavior. Finally, impurity-induced phenomena are highly relevant to engineering lattices of defects on a surface in order to design systems with a Hamiltonian of interest \cite{Ojanen2017,Kaladzhyan2016,Sahlberg2017,Hsieh2017}.

In this paper we study the RKKY interaction between two magnetic impurities localized on a surface of a Dirac semimetal. The surface is chosen to be parallel to the Dirac node separation axis. 
We derive the exchange interaction between two localized spins,  separated by a distance $r$, taking into account  both the bulk and the surface contributions. 
We find that the surface term provides $1/r$
or $1/r^4$ asymptotic scaling behaviors for a linear dispersive Fermi arc in the situation where the spins are separated normal or parallel to the node separation axis, respectively, 
and may thus represent the dominating contribution compared to $1/r^3$ (or $1/r^5$ at the charge neutrality point) evanescence of the bulk term. 

We also demonstrate that there is a salient anisotropy of the RKKY interaction on the surface stemming from a strong anisotropy in the dispersion of the surface states, and most paramountly, from the bulk anisotropic dispersion. Namely, the energy of the interaction decays as $1/r$ 
in the direction perpendicular to the node separation axis, whereas it has a $1/r^4$ power-law decay and an oscillatory behavior in the direction parallel to it, with a period proportional to inverse separation of the Dirac nodes in momentum space $(2k_0)^{-1}$.

Furthermore, we discuss consequences of Fermi arc bending, as well as of a superconducting spin-mixing term lifting the spin degeneracy inherent to Fermi arc surface states in Dirac materials. In particular, the quadratic corrections to the dispersion of Fermi arcs might increase the spatial decay of the surface state contribution to the RKKY interaction.

The paper is organized as follows: In Sec.~\ref{II} we present the model Hamiltonians. In Sec.~\ref{III} we calculate the surface and the bulk contributions to the RKKY interaction of two spins deposited on the surface of a Dirac semimetal, and provide an analysis of the results. 
We present the discussion and conclusions in Secs.~\ref{V} and ~\ref{VI}, respectively, leaving the details of derivations and calculations to the Appendices.

\section{Model}\label{II}

Though our results will apply to most topological Dirac and Weyl semimetals, we have in mind Na$_3$Bi and Cd$_3$As$_2$, two prominent realizations of Dirac semimetals. In these materials the Dirac cones  
 form near the $\Gamma$ point from the pair of $s$-orbitals with $j_z=\pm 1/2$ and  the pair of $p-$orbitals with $j_z=\pm 3/2$ \cite{Wang2013,Cano2017}.
The $k \cdot p$ Hamiltonian for these Dirac semimetals can thus be written as \cite{Cano2017}
\begin{equation}
\mathcal{H}_b(\bs{k}) = -M(k_z) \tau_z  + v (k_x \tau_x \sigma_z -k_y \tau_y),
\label{Hbulk}
\end{equation}
where $M(k_z) \equiv m_0 - m_1 k_z^2$ with $m_0, m_1 > 0$, and $v$ is the Fermi velocity in the plane normal to the $k_z$ axis.  The Pauli matrices $\tau_{x,y,z}$ and $\sigma_{x,y,z}$ act in orbital and spin subspaces, respectively. Hereinafter we set $\hbar=k_B=1$.

The eigenvalues of the Hamiltonian in Eq. (\ref{Hbulk}) determining the bulk spectrum of particles are given by $E_{b,\pm}(\bs{k}) = \pm \sqrt{M^2(k_z) + v^2(k_x^2 + k_y^2)}$, with two Weyl nodes located at $\bs{k}_{\pm} =(0,\,0,\,\pm k_0)$, where $k_0 \equiv \sqrt{m_0/m_1}$. It is worth mentioning that these nodes represent band touching points where $E_{b,\pm}(\bs{k}_\pm)=0$. We note that spectra of Dirac semimetals are spin-degenerate, therefore, each cone contains Weyl fermions with both spin up and spin down. At this stage we make an approximation and deliberately choose not to include any spin-mixing terms into consideration. We leave the discussion of the effects of such terms to Sec.~\ref{V}. 

\begin{figure}	\includegraphics[width=0.9\columnwidth]{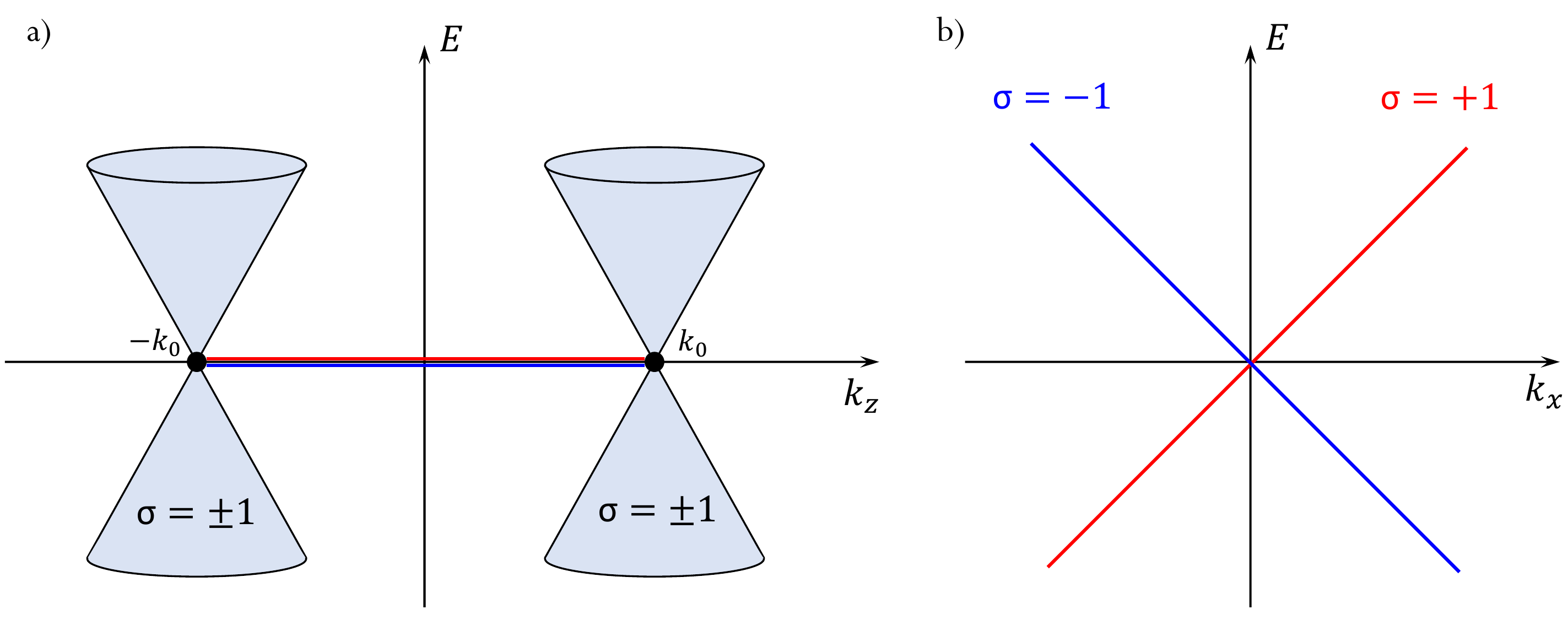}
\caption{(a) Spin-degenerate Dirac cones, and spin-up (spin-down) Fermi arcs at zero energy shown in red (blue). (b) Chiral dispersion of spin-up and spin-down Fermi arcs in $k_x$.}
\label{figspectrum}
\end{figure}

In what follows we consider a Dirac semimetal in the half-space $y \geqslant 0$ with a surface at $y=0$. Fermi arc surface states of the model Hamiltonian in Eq.~(\ref{Hbulk}) can be derived by replacing $k_y \to -i\partial_y$ and assuming that there is no current flowing through the surface. Alternative derivation requires us to introduce at $y<0$ a large-gap insulator circumscribed by the same model as in Eq.~(\ref{Hbulk}), but with $m_0 \to - \infty$ \cite{Okugawa2014}. Both aforementioned ways of treating the boundary can be summarized in the following boundary conditions 
$$
\psi_\sigma(\mathbf{r})\big|_{y=0} \propto \bpm 1\negthickspace+\negthickspace\sigma, 1\negthickspace-\negthickspace\sigma, 1\negthickspace+\negthickspace\sigma, 1\negthickspace-\negthickspace\sigma \epm^\mathrm{T},
$$
where $\sigma = \pm 1$ stand for up and down spins, respectively. Due to spin degeneracy of the bulk Dirac cones, there are two Fermi arcs, one of each spin (see Fig.~\ref{figspectrum}). The solution to the eigenvalue problem, which is determined for $k_z \in \left( -k_0,\,k_0\right)$, gives the wave function of the surface states for $y\geqslant 0$
\begin{equation}
\psi_\sigma(\mathbf{r}) \propto e^{i k_x x} e^{i k_z z} e^{-\frac{M(k_z)}{v}y} \bpm 1\negthickspace+\negthickspace\sigma, 1\negthickspace-\negthickspace\sigma, 1\negthickspace+\negthickspace\sigma, 1\negthickspace-\negthickspace\sigma \epm^\mathrm{T}
\label{SSWF}
\end{equation}
and the eigenvalues $E_{\sigma}=\sigma vk_x$. It is instructive to note that by considering the plane $y=0$ we obtain the largest possible separation between the Fermi arc bound points along the $k_z$ axis in momentum space $2k_0$, whereas  there are no Fermi arcs bound to the surface $z=0$.  The Hamiltonian describing the surface states can be written as
\begin{equation}
\mathcal{H}_s(\bs{k}) = v k_x \sigma_z,
\label{Hsurf}
\end{equation}
where the spin-mixing terms due to surface potential and bulk spin-flip processes are neglected. We note that the surface states obtained above do not disperse in $k_z$, since we have chosen the simplest possible description of the Fermi-arc states allowing us to proceed with our calculations. In the most general case, the choice of boundary conditions \cite{Hashimoto2016}, surface potentials and magnetic fields \cite{Tchoumakov2017}, as well as internode scattering \cite{Devizorova2017} modify the shape of Fermi arcs in momentum space. We provide a brief discussion of the particular case of parabolic Fermi arcs in Sec. \ref{V}, while considering line Fermi arcs in what follows. 

Below we introduce two spins $\bs{S}_i = \left\{ S_{ix},\,S_{iy},\,S_{iz} \right\}$, $i = 1,2$ localized at $\bs{r}_1$ and $\bs{r}_2$, respectively. In order to write down an exchange Hamiltonian we follow Ref.~[\onlinecite{Cano2017}] and hence leave the detailed derivation to Appendix \ref{AppSpinRepresentation}. The structure of the exchange interaction is similar to the Zeeman term with magnetic moments of localized spins playing the role of the magnetic field. As pointed out earlier, the low-energy $k \cdot p$ model is written in the basis of $s$ and $p$ orbitals. The form of the exchange Hamiltonian for $\mathrm{Cd_3As_2}$ is taken from Ref.~[\onlinecite{Cano2017}] and adapted to our basis. Since the basis contains different orbitals we assume that the Zeeman field couples to these bands with different $g$ factors. Therefore, we can formally write:
\begin{align}
\nonumber \mathcal{H}_{b}^{\mathrm{ex}}(\mathbf{r}) = \sum\limits_{i=1,2} \Big[ J_\perp (1 + \tau_z)(\sigma_x S_{ix} + \sigma_y S_{iy}) + \phantom{aaaaa} \\
 (J_z + \delta J_z \tau_z)\sigma_z S_{iz} \Big]
 \delta(\bs{r}-\bs{r}_i)
\label{Hexbulk}
\end{align}
where $J_\perp, J_z$ and $\delta J_z$ are exchange interaction constants, which are generally band-dependent. There is a clear asymmetry in the Hamiltonian in Eq.~(\ref{Hexbulk}) between transverse and $z$ directions originating from the fact that the $z$ axis corresponds to the node separation axis in Eq.~(\ref{Hbulk}). To obtain a similar expression for the surface Hamiltonian, we consider the exchange Hamiltonian in the basis of the Fermi arc surface states defined by Eq.~(\ref{SSWF})
\begin{equation}
\mathcal{H}_{s}^{\mathrm{ex}}(\mathbf{r}) = \sum\limits_{i=1,2} \Big[\sigma_x J_x S_{ix} + \sigma_{y} J_y S_{iy} + J_z\sigma_z S_{iz} \Big]
 \delta(\bs{r}-\bs{r}_i).
\label{Hexsurf}
\end{equation}
Generally, the exchange interaction might be sensitive to possible band bending near the boundary of the crystal, which will be neglected here.

\section{RKKY interaction: normal case}\label{III}

To proceed, we introduce two magnetic impurities with spins $\bs{S}_1$ and $\bs{S}_2$ localized at positions $\bs{r}_1$ and $\bs{r}_2$ at the $y=0$ surface of the Dirac semimetal described by Eq.~(\ref{Hbulk}) in the half-space $y \geqslant 0$, and study their exchange interaction mediated by both the bulk and the surface electrons, using  Eqs.~(\ref{Hbulk}), (\ref{Hexbulk}) and (\ref{Hsurf}), (\ref{Hexsurf}), respectively. \\

\textit{Surface contribution.}
To compute the RKKY interaction we exploit the Matsubara Green function 
in a mixed coordinate-frequency representation. For the surface state at $y=0$, we obtain 
\begin{eqnarray}
\nonumber G_s(\omega_n,\bs{r}) =\int\negthickspace\frac{d \bs{k}}{\left( 2\pi \right)^2 } \frac{i\omega_n +\mu + v k_x \sigma_z}{\left( i\omega_n + \mu\right)^2 -v^2 k_x^2} \times \phantom{aaaaaa} \\
\times \frac{1}{v} M(k_z) \Theta\left[M(k_z) \right] e^{i \bs{k r}},~~~~
\end{eqnarray}
where $\mu$ is the chemical potential, $\bs{k} = (k_x, k_z)$, $\bs{r} = (x,z)$, and $\omega_n \equiv \pi T (2n+1), n \in \mathbb{Z}$ being the fermionic Matsubara frequencies. The Heaviside step function ensures that the surface states exist only for $k_z \in \left( -k_0,\,k_0\right)$, whereas the factor $M(k_z)/v$ stems from normalization of the surface states. Using Jordan's lemma one finds
\begin{align}
\nonumber G_s(\omega_n,\bs{r}) = -\frac{i}{2v} \frac{2m_1\left( \sin k_0 z - k_0 z \cos k_0 z\right) }{\pi v z^3}\times \\
 \times \left(\sgn \omega_n\, + \sigma_z \sgn x\, \right) e^{\left(i\mu -\omega_n \right)\frac{|x|}{v}\sgn \omega_n }.
\label{GsurfFcoord}
\end{align}
Note that the Green function is zero at $k_0 =0$ due to vanishing of the Fermi arc.
The RKKY interaction $\mathcal{E}\left(\bs{r} \right) = \mathcal{E}_s\left(\bs{r} \right)+\mathcal{E}_b\left(\bs{r} \right)$, where $\bs{r} \equiv \bs{r}_1 - \bs{r}_2 \equiv \left(x_1 - x_2 ,\,z_1 - z_2 \right)$ and $\mathcal{E}_{s,b}\left(\bs{r} \right)$ are the surface and bulk contributions, respectively, can be calculated perturbatively in powers of the exchange coupling constants. At the lowest order, the RKKY energy is given by the spin susceptibility diagram \cite{Abrikosov1988} which can be expressed in terms
of the following summation over the Matsubara frequencies
\begin{eqnarray}
\mathcal{E}_s \left(\bs{r} \right) = T\negthickspace\sum\limits_{i,j; n} \negthickspace\tr \left[ J_i S_{1i} \sigma_i G_s (\omega_n,\bs{r}) J_j S_{2j} \sigma_j G_s(\omega_n,-\bs{r})\right],~~~~
\label{Bubblediagram}
\end{eqnarray}
with $i,j \in \left\{x,y,z\right\}$. Performing the summations, we get
\begin{align}
\nonumber \mathcal{E}_s\left(\bs{r} \right) &=  - \frac{1}{v^2} \frac{4m^2_1\left( \sin k_0 z - k_0 z \cos k_0 z\right)^2}{\pi^2 v^2 z^6} \frac{T}{\sh \frac{2\pi |x|T}{v}} \times \phantom{aaaaaaaaaaa}\\ 
\nonumber &\times \left[\left(J_x^2 S_{1x}S_{2x}+J_y^2 S_{1y} S_{2y} \right) \cos \frac{2 x \mu}{v} \right. \\
&\left. \phantom{aaaaaa}+ J_x J_y \left( S_{1x}S_{2y} - S_{1y}S_{2x}\right) \sin \frac{2 x\mu}{v} \right].
\label{RKKYsurface} 
\end{align}
This expression is one of our main results that we analyze in detail below.

At low chemical potentials, such that $|\mu x|/v<1$, the RKKY interaction has a negative sign, and thus favors a ferromagnetic alignment of spins. This result is reminiscent of the interaction between two magnetic impurities on the surface  of a topological insulator \cite{Liu2009,Biswas2010,Garate2010,Zhu2011,Abanin2011,Tokura2012,Zyuzin2014, Pankratov_RKKY}.
 
Provided the separation between the localized spins along the $z$ axis is much larger than the inverse separation between the Dirac nodes 
$\propto (2k_0)^{-1}$, the RKKY energy vanishes due to the factor $1/z^4$ with oscillations given by $\cos 2k_0 z$. At low temperatures, in the limit of large separation along the $x$ axis, the interaction decays as $1/|x|$ in the $x$ direction. This result illustrates the high anisotropy of the RKKY interaction. Namely, at low temperature, the interaction mainly occurs through two impurities which are quasialigned along the $x$ direction. This is a direct consequence of the fact that the Dirac nodes are aligned along $k_z$, while the anisotropic spectrum of the Fermi arc states being dispersionless in $k_z$ and having a chiral dispersion in $k_x$ is responsible for the $1/|x|$ decay in the $x$ direction. 

Such peculiar distinction between $x$ and $z$ directions can be potentially used to engineer impurity lattices with anisotropic hopping constants. Indeed, if we consider a set of dilute magnetic impurities on top of this surface, they will only interact along stripes or bands parallel to the $x$ axis, the width of the stripe being given by $k_0^{-1}$, while having much weaker exchange interaction in the $z$ direction. In the general case, the direction of the stripe will be given by the direction orthogonal to the node separation line, even if the Fermi arcs disperse in both $k_x$ and $k_z$ on the surface.

Another important issue to note is the presence of the 
Dzyaloshinskii-Moriya term in Eq.~(\ref{RKKYsurface}) proportional to $\left[\bs{S}_1 \times \bs{S}_2 \right]_z$. Such terms are usually found in systems with extrinsic or intrinsic spin-orbit coupling and/or spin-valley coupling. Some of the examples include low-dimensional systems such as Rashba 2DEG and nanowires, surfaces of the topological insulators, and carbon nanotubes. 
In the present case, the surface of the Dirac semimetal hosts two counter propagating Fermi arc states with opposite spins, which can be mapped to the chiral edge states of a two-dimensional topological insulator. Although, the bulk is inversion symmetric, the Dzyaloshinskii-Moriya  term is expected to be present at the surface due to the interarc backscattering at finite $\mu$.

\textit{Bulk contribution.}
To find the bulk contribution to RKKY interaction, we first compute the bulk Green's function
\begin{align}
\nonumber G_b(\omega_n,\bs{r}) = \phantom{aaaaaaaaaaaaaaaaaaaaaaaaaaaaaaaaaaaa} \\
\negthickspace\int\negthickspace\frac{d \bs{k}}{\left( 2\pi \right)^3 } \frac{i\omega_n +\mu - M(k_z) \tau_z + v(k_x \tau_x \sigma_z - k_y \tau_y)}{\left(i\omega_n + \mu\right)^2 -M^2(k_z) - v^2(k_x^2+k_y^2 ) } e^{i \bs{k r}}.
\label{bulkG}
\end{align}
A detailed calculation of this integral is given in Appendix \ref{AppBulkGFcalculation}. Above, $\bs{k} = \left\{k_x,\,k_y,\,k_z \right\}$ and  $\bs{r} \equiv \bs{r_1 - r_2} \equiv \left(x_1 - x_2,\,y_1 - y_2,\,z_1 - z_2 \right)$.
Finally, we compute the bulk contribution into RKKY interaction using an analog of Eq.~(\ref{Bubblediagram}) for the bulk case
\begin{equation}
\mathcal{E}_b\left(\bs{r} \right) = T \sum\limits_{i,j;n} \tr \left[ V_{1i} G_b (\omega_n,\bs{r}) V_{2j} G_b (\omega_n,-\bs{r})\right],
\label{BubblediagramBulk}
\end{equation}
where $i,j \in \left\{x,y,z\right\}$. From Eq.~(\ref{Hexbulk}) we define $V_{ai}$ as 
$V_{ax} \equiv J_\perp (1+\tau_z)\sigma_x S_{ax}$,
$V_{ay} \equiv J_\perp (1+\tau_z)\sigma_y S_{ay}$, and 
$V_{az} \equiv (J_z+\delta J_z \tau_z)\sigma_z S_{az}$
with $a \in \left\{1,2\right\}$ referring to $\bs{S}_1$ or $\bs{S}_2$. Using the fact that the localized spins are on the surface $y=0$, we set $y=y'=0$ in Eq.~(\ref{BubblediagramBulk}) and, summing up over $i$, $j$, and Matsubara frequencies, we get the final result 

\begin{align}
 \mathcal{E}_b(\bs{r}) = -C_1 J_\perp^2 \left( S_{1x}S_{2x} + S_{1y}S_{2y} \right)- C_2 J_z^2 S_{1z}S_{2z},
\label{RKKYbulk}
\end{align}
where exact expressions for $C_{1,2}$, which can be found in Appendix \ref{AppMatsubaraFrequencySummation}, are quite complicated in the most general case. For the sake of clarity we present here their asymptotic behavior in the limit of $\mu \to 0$ and $T \to 0$, where we find
\begin{equation}
C_{1} \propto |x|^{-5},\;C_{2} \propto 2\left(1-\frac{3\delta J_z^2}{2J_z^2} \right)|x|^{-5},
\end{equation}
if magnetic impurities are aligned along the $x$ axis, \textit{i.e.} $z=0$, and 
\begin{equation}
C_{1,2} \propto - |z|^{-5} \left[1 - \beta \cos 2k_0 z \right],
\end{equation}
where $\beta > 0$ for Cd$_3$As$_2$, if magnetic impurities are aligned along the $z$ axis, \textit{i.e.} $x=0$. The dominating terms in the bulk contribution decay as $1/|z|^5$ at large distances and oscillate with a frequency determined by the cone separation in momentum space $2k_0$. Such a behavior is in agreement with previous studies about the RKKY interaction in a bulk 3D semimetal \cite{Sun2015,Chang2015,Hosseini2015}. However, in the doped semimetal with $\mu \neq 0$ one has a standard power-law decay inherent in three dimensions
\begin{equation}
C_{1} \propto - |x|^{-3} \cos \frac{2\mu x}{v},\; C_2 \propto \frac{\delta J_z^2}{J_z^2} C_1
\end{equation}
if spins are separated in the $x$ axis, and 
\begin{align}
C_{1,2} &\propto - |z|^{-3} \bigg[ \sin 2k_0 z \sin\frac{2\mu z}{v}  \\\nonumber
&+\left(\beta_{1,2} \sin^2 k_0 z - \gamma_{1,2}\cos^2 k_0 z\right)\cos\frac{2\mu z}{v} \bigg],
\end{align}
if magnetic impurities are aligned along the $z$ axis, where the signs of coefficients $\beta_{i}, \gamma_i$ depend on the chemical potential $\mu$. Therefore, at a nonzero value of the chemical potential $\mu$ the bulk contribution decays as $1/r^3$.

\begin{figure}	\includegraphics[width=0.43\columnwidth]{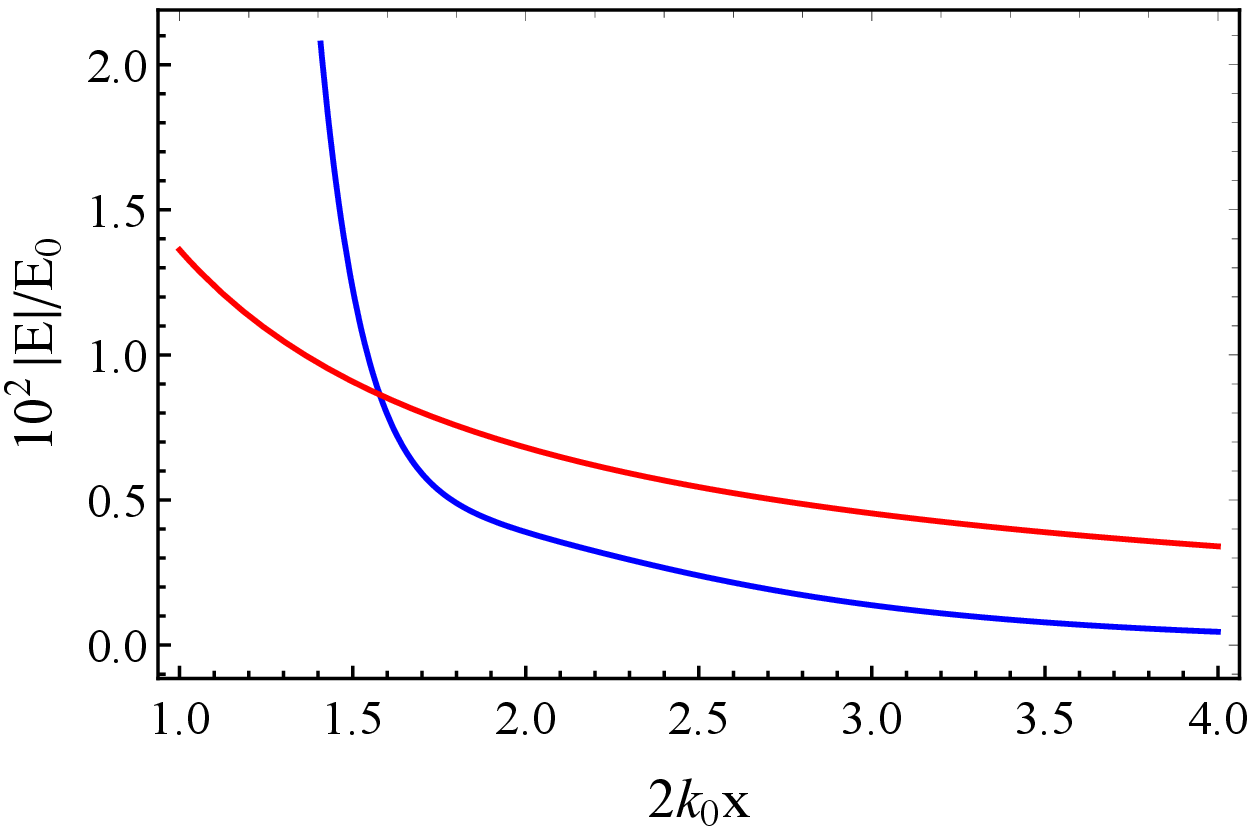}
\includegraphics[width=0.45\columnwidth]{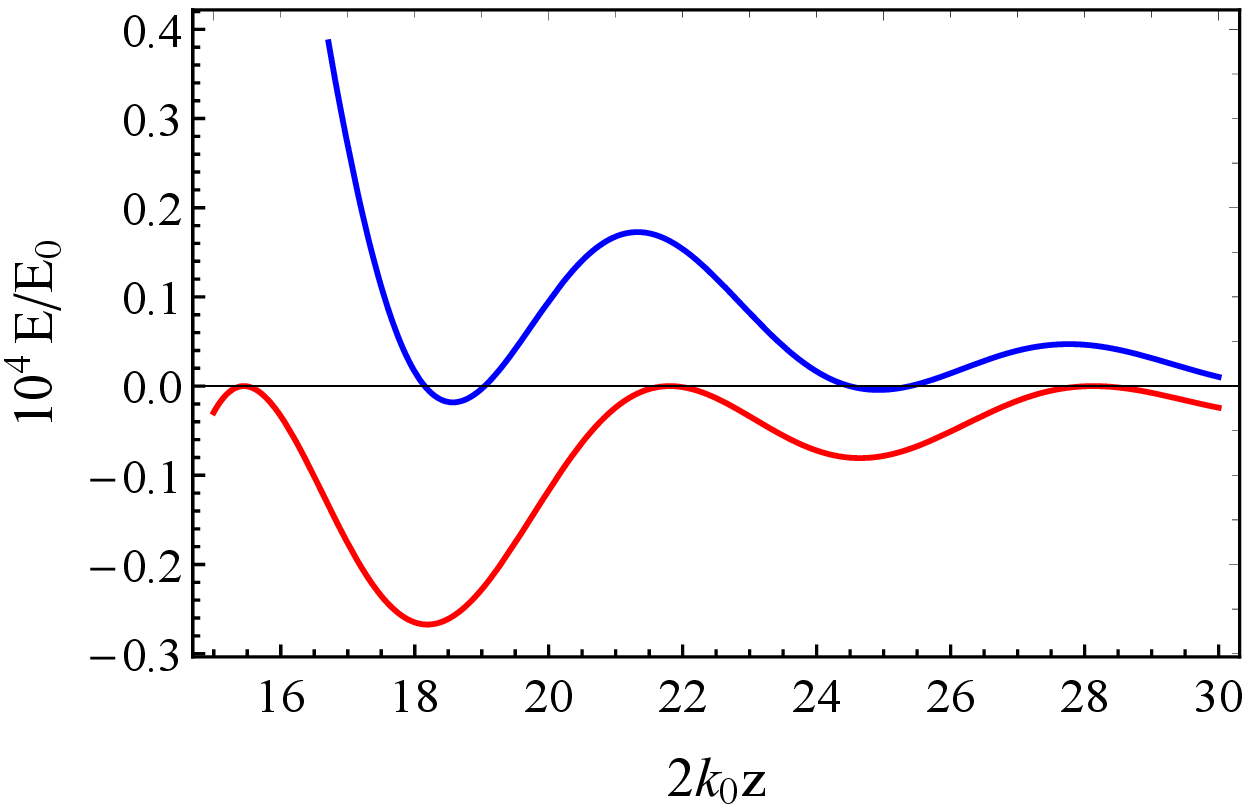}\\
\includegraphics[width=0.45\columnwidth]{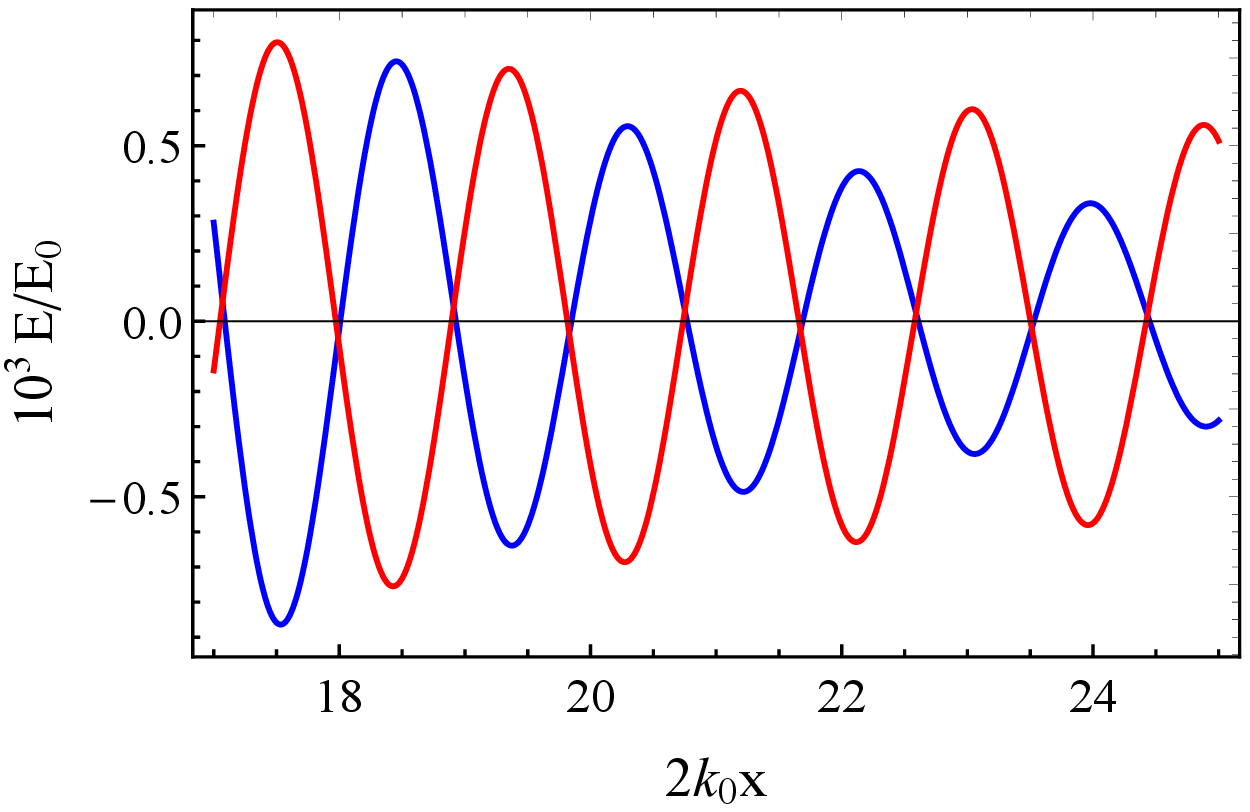}
\includegraphics[width=0.45\columnwidth]{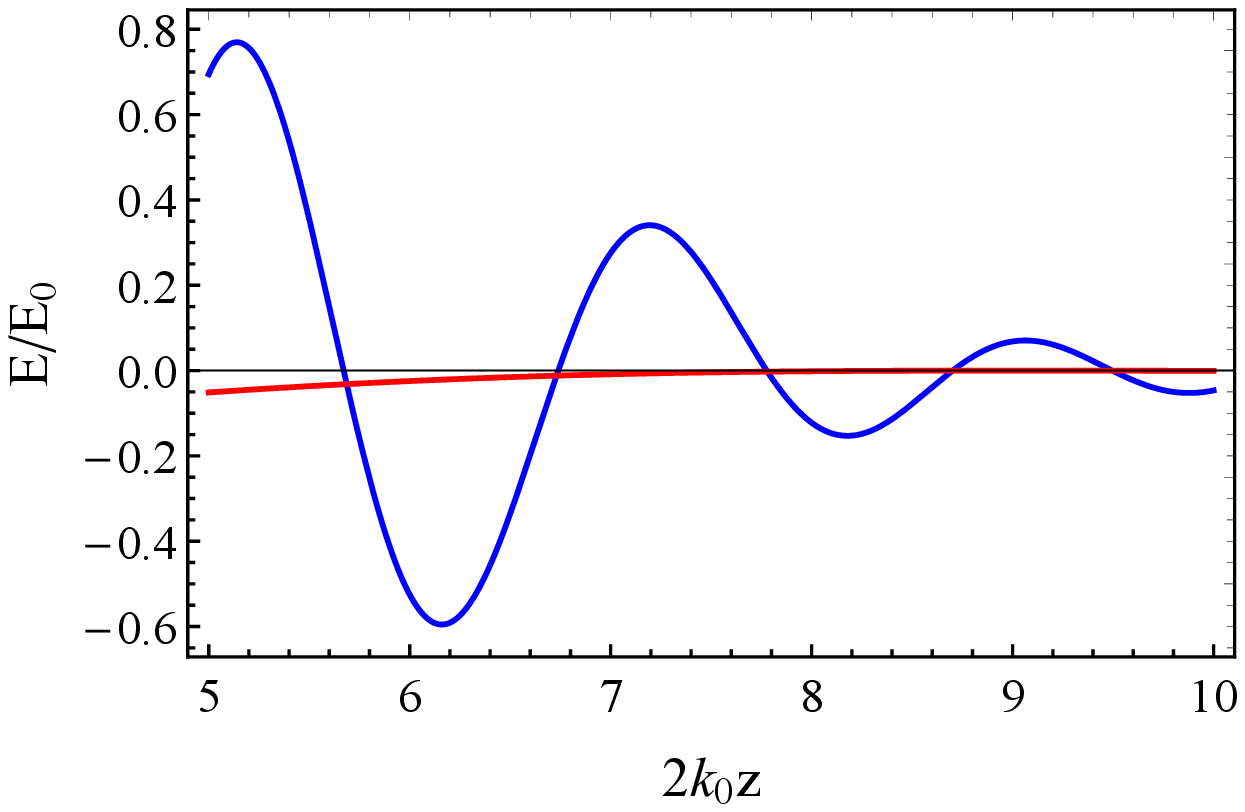}\\
	\caption{Surface (red curves) and bulk (blue curves) contributions to the RKKY interaction normalized to a characteristic energy scale $E_0 \equiv J^2 k_0^5/v$ plotted as a function of the dimensionless distances between the localized spins along the $x$ axis (left column) and $z$ axis (right column). We take $\mu = 0$ and $\mu = 0.1\,$eV for upper and lower panels correspondingly. We assume the distance between the impurities to be $z=10\,$\AA\; (upper-left panel), $z=1\,$\AA\; (lower-left panel), and $x=10\,$\AA\; (upper-right panel), $x=1\,$\AA\; (lower-right panel). We set: $S_{1i}=S_{2i}=1$, $J_{x,y}=J_\perp = J, J_z=\delta J_z = J$, where we set the exchange coupling constant is measured in eV$\cdot$\AA$^3$, and we take realistic parameters for Cd$_3$As$_2$ given by Eq.~(S45) from Ref.~[\onlinecite{Cano2017}]: $v=0.889\,$eV$\cdot$\AA, $k_0 = 0.033\,\text{\AA}^{-1}$, $m_1=18.77\,$eV$\cdot$\AA$^2$, and $\gamma = 0.513$.}
	\label{FigBulkvsSurface}
\end{figure}

\textit{Bulk versus surface.} 
To analyze the full RKKY interaction of two magnetic impurities localized on the surface of a Dirac semimetal we need to take into account both the bulk and the surface contributions. We assume that they can be taken into account independently \cite{Gorbar2016}, and consider the simplest case of $T \to 0$. In Fig.~\ref{FigBulkvsSurface} we plot separately the surface and the bulk contributions given by Eqs.~(\ref{RKKYsurface}) and (\ref{RKKYbulk}) taken for various distances between the localized spins at two different values of the chemical potential: $\mu = 0$ (the charge neutrality point) and $\mu = 0.1\,$eV. For $\mu=0$ along the $x$ axis (the upper-left panel) the surface contribution dominates the bulk one with $1/|x|$ asymptotic behavior versus $1/|x|^5$. Similarly, along the $z$ axis the surface contribution mostly dominates the bulk one with a $1/z^4$ power-law evanescence versus $1/|z|^5$, respectively (upper-right panel), except for the vicinity of the points where $\sin k_0 z - k_0 z \cos k_0 z = 0$. However, the situation is getting more complicated when we consider $\mu \neq 0$ (lower row in Fig.~\ref{FigBulkvsSurface}), since in that case the bulk contribution mostly dominates the surface one along the $z$ axis, whereas it is vice versa along the $x$ axis.

\begin{figure}	\includegraphics[width=0.7\columnwidth]{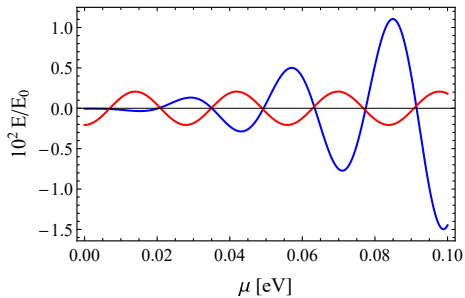}
	\caption{Surface (red curves) and bulk (blue curves) contributions to the RKKY interaction normalized to a characteristic energy scale $E_0 \equiv J^2 k_0^5/v$ plotted as a function of the chemical potential $\mu$. We assume the distance between the impurities to be $z=10\,$\AA\;   along the $z$ axis and $x=100\,$\AA\; along the $x$ axis. Other parameters are taken to be the same as in Fig.~\ref{FigBulkvsSurface}.}
	\label{FigBulkvsSurfacemudep}
\end{figure}

It is also worth discussing the chemical potential dependence of the RKKY interaction. Thus in Fig.~\ref{FigBulkvsSurfacemudep} we show the aforementioned dependence for both the bulk and the surface terms using the same color code as in Fig.~\ref{FigBulkvsSurface}. It is clear that both contributions have oscillatory behavior in $\mu$, therefore, depending on the value of the chemical potential we can have either the bulk or the surface contribution dominating in a given direction at a given distance.


\section{Discussion}\label{V}

One of in the important questions is the stability of the Fermi arc surface states with respect to the scattering on magnetic impurities.

In the case of the ferromagnetic Weyl semimetal the unidirectional surface states contribute to RKKY only by interfering with the bulk states \cite{Duan2018}. The spectrum of the Fermi arc is protected from the gap opening, provided that the electron spin splitting induced by the magnetic impurities does not merge the Weyl points.

The situation is different in the case of Dirac semimetals discussed in this paper. Single magnetic impurity allows for backscattering between two counterpropagating Fermi arc surface states with opposite spins described by Hamiltonian in Eq. \ref{Hsurf}. Many random impurities placed on the surface of the semimetal, provided that the separation between them is less than $v/|\mu|$ and larger than the localization length of the surface states, tend to align collinear with respect to each other, which leads to a gap opening in the Fermi arc states spectrum. This result is similar to what one expects for the helical surface states of topological insulator \cite{Liu2009,Biswas2010,Garate2010,Zhu2011,Abanin2011}. 

The interference contribution to the RKKY interaction between magnetic impurities at the surface of the system due to electron scattering between the surface and the bulk states was studied for topological insulator in Ref. \cite{KKBurmistrov} and for ferromagnetic noncentrosymmetric Weyl semimetal in Ref. \cite{Duan2018}. This correction vanishes in the case of topological insulator provided $\mu \rightarrow 0$, while it dominates the RKKY interaction in the case of a Weyl semimetal. Here we show that the interference correction can be small compared to the purely surface contribution.

Finally, in the case of graphene, the sign and the amplitude of the RKKY interaction depends strongly on the sublattice position of impurities, favoring antiferromagnetic spin ordering of many randomly placed impurities \cite{Saremi2007,Hwang2008,Black-Schaffer2010,Sherafati2011,Kogan2011}. 
The back action effect of spin ordering might as well lead to the gap in the electron spectrum \cite{Braunecker2009prl}.

In the previous sections, we have calculated both the surface and bulk contributions to the RKKY interactions between two impurities.
Before discussing our results,  we would like to point out that the preferred ground state  of  interacting localized magnetic impurities is nonuniversal and does not simply depend on the RKKY interaction but might depend as well on crystal symmetries, spin-orbit coupling and exchange interaction parameters.

We have used Cd$_3$As$_2$ as an experimentally relevant example of a Dirac semimetal,
where our results can be potentially tested. Despite employing the model and parameters for a specific Dirac semimetal Cd$_3$As$_2$,
the spatial dependence of the RKKY interaction, which depends on the Friedel oscillations of the electron wavefunction, has a universal form and similar results are expected for 
other topological Dirac-Weyl semimetals (for example, two-dimensional states in band-inverting heterojunctions \cite{Pankratov_Review}, three-dimensional TaAs semimetal, or HgTe and half-Heuslers compounds under strain \cite{Weyl_review}).


It is worth noting that we neglected spin-mixing terms while considering the normal case. Such terms may arise due to natural reasons (such as \textit{e.g.} imperfections on the surface), or in the presence of proximity-induced superconducting pairing $\Delta$ on the surface, as discussed in Appendix \ref{AppRKKYSC}. In the latter case we have obtained an additional term $\propto \Delta S_{1z}S_{2z}$, whereas the former would introduce a competing contribution $\propto -\Delta_{mix} S_{1z}S_{2z}$, otherwise absent in Eq.~(\ref{RKKYsurface}).

Following Ref.~[\onlinecite{Gorbar2016}] we have assumed that bulk and surface contributions to the RKKY interaction can be taken into account separately. This is a strong assumption, noting that the surface states of Dirac semimetals are not protected by a bulk gap. In this scheme, we have basically neglected any surface-bulk interaction terms that can participate in the RKKY interaction. 
However, the fact that the surface contribution provides a $1/|x|$ leading behavior compared to the $1/|x|^5$ (or $1/|x|^3$) decrements of the bulk makes us confident that the results we have obtained may survive this approximation. 
To go beyond that argument, we qualitatively evaluate the account of a surface-bulk mixing interference  term to the RKKY exchange energy in Appendix \ref{AppBulkSurfaceInterference} and show that it indeed decays as $1/r^6$ ($1/r^5$) for $\mu=0$ ($\mu \neq 0$), and can thus be  neglected.

It is also important to compare the strength of the exchange interaction with the dipole-dipole interaction between magnetic impurities, which can be described by the Hamiltonian
$H_{d-d}(\mathbf{r}) = (g\mu_N)^2[\mathbf{S}_1\cdot\mathbf{S}_2 -3(\mathbf{S}_1\cdot \hat{\mathbf{r}})(\mathbf{S}_2\cdot \hat{\mathbf{r}})]/2r^3 $, where $\mu_N$ is the magneton, $\hat{\mathbf{r}}$ is the unit vector, and $g$ is the $g$ factor of the magnetic impurity.
At distances between the localized spins $r\ll J/g\mu_N$ the coupling might be dominated by the dipole-dipole interaction, which holds as long as we neglect the magnetocrystalline anisotropy.

Finally, it is instructive to discuss the effects of Fermi arc bending. In order to do so rigorously it is required to study a model that embodies curved Fermi arcs. The existing models employ sophisticated mechanisms such as, e.g., internode scattering, phenomenological boundary conditions, or interface potentials with  magnetic fields. Thus, to date, to our knowledge, there exist no general models yielding curved Fermi arcs. Therefore,  we have decided to follow a different strategy and to qualitatively analyze the effect of a Fermi-arc bending. Below we focus on evaluating how this change of the Fermi-arc dispersion is able to modify the power-law exponent of the spatial dependence of the RKKY interaction. We note that finding a general model yielding curved Fermi arcs and rigorously analyzing the effect of Fermi arc bending on the RKKY interaction lie beyond the scope of this work. For the sake of simplicity, we assume that the Fermi arc is parabolic, i.e. the surface Hamiltonian in Eq.~(\ref{Hsurf}) reads
\begin{equation}
\mathcal{H}_{s} = \left[ v k_x - \gamma\left(k_z^2-k_0^2 \right)\right] \sigma_z,
\end{equation}
where $\gamma \neq 0$ is responsible for the Fermi arc curvature. In the presence of such a curved Fermi arc 
we obtain the final expression for the surface contribution by replacing in Eq.~(\ref{RKKYsurface})
\begin{align}
\nonumber\frac{4m^2_1\left( \sin k_0 z - k_0 z \cos k_0 z\right)^2 }{\pi^2 v^2 z^6}  \to \phantom{aaaaaaaaaaaaaaa}\\
\left|\; \int\limits_{-k_0}^{k_0} \negthickspace \frac{dk_z}{2\pi} \frac{M(k_z)}{v}e^{i k_z z} e^{i \frac{\gamma}{v}x (k_z^2-k_0^2)} \right|^2.
\end{align}
The integral above can be done in terms of special functions, however, it is worth discussing the limit of small Fermi arc curvature in which we can expand it as a Taylor series as follows:
\begin{align}
\left|\; \int\limits_{-k_0}^{k_0} \negthickspace \frac{dk_z}{2\pi} \frac{M(k_z)}{v}e^{i k_z z} e^{i \frac{\gamma}{v}x (k_z^2-k_0^2)} \right|^2 \sim \phantom{aaaaaaaaaaaaaaaa}\\
\nonumber\frac{4m^2_1\left( \sin k_0 z - k_0 z \cos k_0 z\right)^2 }{\pi^2 v^2 z^6} + \phantom{aaaaaaaaaaaaaaaaaaaaa} \\
\nonumber\frac{64m_1^2 \gamma^2 \left( k_0^2z^2 \sin k_0 z + 3 k_0 z \cos k_0 z - 3 \sin k_0 z \right)^2}{\pi^2 v^2 z^{10}} x^2.
\end{align}
At $k_0 |x| \ll 1$ and/or $\gamma k_0 /v \ll 1$ the integral above simplifies to the expression we had before, because physically this corresponds to a very flat Fermi arc, thus reproducing the case of a line Fermi arc we have considered in Sec.~\ref{III}. It is easy to verify numerically that at large distances $k_0 |x| \gg 1$ the surface contribution acquires an additional factor of $1/x$, therefore, making the overall $x$ dependence to be $\propto 1/x^2$. As before, the $z$ dependence of the factor computed above is oscillatory and has a leading power-law decay of $1/z^4$. It is clear that, first, the shape of Fermi arcs affects the power-law decay of the RKKY interaction on the surface, and second, it is worth noting that the salient anisotropy between $x$ and $z$ directions persists because it is intrinsically woven into the model with the Dirac nodes along $k_z$.

\section{Conclusions}\label{VI}

We have analyzed the RKKY interaction between magnetic impurities placed on a surface of a Dirac semimetal hosting Fermi arc surface states. We have demonstrated that the Fermi arc contribution may be dominating in the RKKY interaction energy due to a weaker power-law decay compared with the bulk contribution.

Moreover, due to strong momentum-space anisotropy in the spectrum of the bulk states (with two Dirac nodes being aligned along the $k_z$ axis in momentum space and separated by $2k_0$) there is a salient anisotropy of the RKKY interaction in real space. In the $x$ direction, it decays as $1/|x|$, whereas in the $z$ direction its behavior is circumscribed by a power-law decay of $1/z^4$ and an oscillatory factor with frequency $2k_0$. These peculiar features can be potentially used for engineering low-dimensional lattices of interacting spins with anisotropic hopping parameters in different directions. We have checked also that this hallmark persists in the presence of parabolic Fermi arcs, with a $1/x^2$ decay in the $x$ direction in that case.

Furthermore, we have studied the effect of spin-mixing terms on the surface by considering the simplest case of a conventional superconducting pairing. We have found that the presence of such terms leads to an additional contribution to the RKKY interaction on the surface proportional to $z$ components of interacting spins and to the amplitude of the spin-mixing term.

While finalizing this work, we have become aware of a paper addressing the problem of RKKY interaction on the surface of a ferromagnetic noncentrosymmetric Weyl semimetal with two Weyl cones in the band structure \cite{Duan2018}. In that model, purely Fermi arc surface states do not contribute to the RKKY interaction. 
\\

\section{Acknowledgements}
V.K. and P.S. were partially supported by French Agence Nationale de la Recherche through the contract ANR Mistral. V.K. would like to acknowledge the ERC Starting Grant No. 679722 and the Roland Gustafsson foundation for theoretical physics, as well as fruitful discussions with Sergue\"i Tchoumakov and Lo\"ic Herviou, and an important comment from Igor Burmistrov. A.A.Z. is supported by the Academy of Finland.

\bibliography{biblio_RKKY_v2}

\newpage
\widetext
\appendix

\section{Spin representation for a Dirac semimetal}\label{Appendix}
\label{AppSpinRepresentation}
In this appendix we show how to introduce correctly a Zeeman field into a Dirac semimetal using the example of Cd$_3$As$_2$. Following Eq. S47 from Ref.~[\onlinecite{Cano2017}] we write down the $k\cdot p$ Hamiltonian in the basis of states with total angular momenta $\{ \left|3/2 \right\rangle, \left|1/2 \right\rangle, \left|-1/2 \right\rangle, \left|-3/2 \right\rangle\}$:
\begin{align}
\mathrm{H}(\mathbf{k}) = 
\left[\begin{array}{cccc}
M(k_z) & v k_{-} & 0 & 0 \\
v k_{+} & -M(k_z) & 0 & 0 \\
0 & 0 & -M(k_z) & -v k_{-} \\
0 & 0 & -v k_{+} & M(k_z)\\
\end{array}\right],
\label{Hbulkdiffbasis}
\end{align} 
where $M(k_z) = m_0 - m_1 k_z^2$ and $k_{\pm} = k_x\pm i k_y$. We note that we kept only the most relevant terms for our problem. According to Eq. S47 the Zeeman term associated with a magnetic field $\bs{B} = \{B_x,\,B_y,\,B_z \}$ reads
\begin{align}
\mathrm{H}_{Z} = g_s
\left[\begin{array}{cccc}
0& 0 & 0 & 0 \\
0& B_z & B_x-i B_y & 0 \\
0 & B_x + i B_y & -B_z & 0 \\
0 & 0 & 0& 0\\
\end{array}\right] + 
 g_p
\left[\begin{array}{cccc}
B_z& 0 & 0 & 0 \\
0& 0 & 0 & 0 \\
0 & 0 & 0 & 0 \\
0 & 0 & 0& -B_z\\
\end{array}\right],
\label{HZeemandiffbasis}
\end{align} 
where we introduced different $g$ factors, namely $g_s$ stands for bands corresponding to $J_z = \pm 1/2$, whereas $g_p$ refers to those with $J_z =\pm 3/2$. Other terms are mixed with the bands further away in energy and assumed to be small.

In order to make the Hamiltonian in Eq.~(\ref{Hbulkdiffbasis}) resemble the one we use in the main text in Eq.~(\ref{Hbulk}), we switch to a new basis $\{ \left|1/2 \right\rangle, \left|-1/2 \right\rangle, \left|3/2 \right\rangle, \left|-3/2 \right\rangle\}$ and we get 
\begin{align*}
\mathcal{H}_b(\mathbf{k}) = 
\left[\begin{array}{cccc}
-M(k_z) & 0 & v k_{+} & 0 \\
0 & -M(k_z) & 0 & -v k_{-} \\
v k_{-} & 0 & M(k_z) & 0 \\
0 & -v k_{+} & 0 & M(k_z)\\
\end{array}\right]
\end{align*} 
and
\begin{align*}
\mathcal{H}_{Z} = g_s
\left[\begin{array}{cccc}
B_z& B_x-i B_y & 0 & 0 \\
B_x+i B_y  & -B_z & 0 & 0 \\
0 & 0 & 0 & 0 \\
0 & 0 & 0& 0\\
\end{array}\right] + 
 g_p
\left[\begin{array}{cccc}
0& 0 & 0 & 0 \\
0& 0 & 0 & 0 \\
0 & 0 & B_z & 0 \\
0 & 0 & 0& -B_z\\
\end{array}\right] 
\end{align*} 
corresponding to Eqs.~(\ref{Hbulkdiffbasis}) and (\ref{HZeemandiffbasis}) written in the new basis. Rewriting them in terms of two sets of Pauli matrices introduced in the main text we obtain
\begin{align}
\mathcal{H}_b(\mathbf{k}) &= -M(k_z) \tau_z + v(k_x\tau_x\sigma_z - k_y \tau_y)\\
\mathcal{H}_Z &= (g_z + \delta g_z \tau_z)\sigma_z B_z + g_{\perp}(\tau_0 + \tau_z)(\sigma_x B_x + \sigma_y B_y)
\end{align}
where $g_z\equiv (g_s+g_p)/2$, $\delta g_z\equiv (g_s-g_p)/2$, $g_{\perp} \equiv g_s/2$.
Since the matrix structure of the exchange interaction is expected to be similar to the Zeeman term we can formally write for a localized spin $\bs{S} = \{S_x,\,S_y,\,S_z \}$:
\begin{align}
\mathcal{H}^{ex}_{b} = J_{\perp}(\tau_0 + \tau_z)(\sigma_x S_x + \sigma_y S_y) + (J_z + \delta J_z \tau_z)\sigma_z S_z
\label{AppHexbulk}
\end{align}

To obtain an analog of this expression for the exchange interaction with the surface states, we project Eq.~(\ref{AppHexbulk}) onto the basis of the surface states defined in Eq.~(\ref{SSWF}) and we get
\begin{align}
\mathcal{H}^{ex}_{s} = J_{\perp} (\sigma_x S_x + \sigma_y S_y) + J_z \sigma_z S_z
\end{align}

\section{Calculation of the bulk Green's function in real space}
\label{AppBulkGFcalculation}
In what follows we compute the real-space form of the bulk Green's function from Eq.~(\ref{bulkG}):
\begin{align}
\nonumber G_b(\omega_n,\bs{r}) = &- \int\frac{d \bs{k}}{\left( 2\pi \right)^3 } \frac{\left( i\omega_n +\mu\right) - M(k_z) \tau_z + v(k_x \tau_x \sigma_z - k_y \tau_y)}{M^2(k_z) + v^2(k_x^2+k_y^2 ) + \left(\omega_n - i \mu\right)^2} \cdot e^{i \bs{k r}} =\\
\nonumber &-\left[ \left( i\omega_n +\mu\right) - (m_0+m_1\partial_z^2) \tau_z + iv \partial_x \tau_x \sigma_z - i v \partial_y \tau_y \right]  I(\bs{r}) \equiv \\
&\phantom{--}I_0(\bs{r})  + I_x(\bs{r}) \tau_x\sigma_z + I_y(\bs{r}) \tau_y + I_z(\bs{r}) \tau_z,
\label{AppbulkG}
\end{align}
where we defined a set of functions $I_0(\bs{r})$ and $I_i(\bs{r})$ ($i \in \left\{x,\,y,\,z \right\}$) expressed in terms of an auxiliary integral 
\begin{align}
I(\bs{r}) = \int\frac{d \bs{k}}{\left( 2\pi \right)^3 } \frac{e^{i \bs{k r}}}{M^2(k_z) + v^2(k_x^2+k_y^2 ) + \left( \omega_n - i \mu\right)^2} 
\label{auxint}
\end{align}
in the following way:
\begin{align}
I_0(\bs{r}) \equiv -\left( i\omega_n +\mu\right) I(\bs{r}),\; I_x(\bs{r}) \equiv iv \partial_x I(\bs{r}),\; I_y(\bs{r}) \equiv -iv \partial_y I(\bs{r}),\; I_z(\bs{r}) \equiv \left( m_0 + m_1 \partial_z^2 \right) I(\bs{r}).
\label{AppdefI0xyz}
\end{align}
Above, $\bs{k} = \left\{k_x,\,k_y,\,k_z \right\}$ and  $\bs{r} \equiv \bs{r_1 - r_2} \equiv \left(x_1 - x_2,\,y_1 - y_2,\,z_1 - z_2 \right)$. A detailed calculation of $I(\bs{r})$ is presented below.\\

Most of the contribution of the integrand in $I(\bs{r})$ is coming from around Weyl points, $k_z = \pm k_0$. Therefore, we can linearize the spectrum around those points and sum up the results of integrations separately. We start by $k_z = k_0$ (the expression for the other Weyl point is obtained along similar lines):
\begin{equation*}
I_+(\bs{r}) = \frac{1}{(2\pi)^3v^2}\int d \bs{k} \frac{e^{i \bs{k r}}}{k_x^2+k_y^2 + \frac{1}{\gamma} \left( k_z - k_0 \right)^2 + \left( \omega_n - i \mu\right)^2/v^2} = \bullet
\end{equation*}
We use spherical coordinates defined as follows:
\begin{align*}
k_x &= k \sin \theta \cos \phi  \\
k_y &= k \sin \theta \sin \phi \\
k_z &= k_0 + \sqrt{\gamma} k \cos \theta,
\end{align*}
where $\gamma \equiv \frac{v^2}{4m_0m_1}$. The Jacobian of such a transformation is $ \sqrt{\gamma}\, k^2 \sin \theta$. Thus we get
\begin{align*}
\bullet = \sqrt{\gamma}\, \frac{e^{i k_0 z}}{(2\pi)^2v^2}\int\limits_{0}^\infty k^2 dk \int\limits_{0}^\pi \sin \theta d\theta \frac{e^{i k \tilde{r} \cos\theta}}{k^2 + \left( \omega_n - i \mu\right)^2/v^2} = \bullet \bullet
\end{align*}
where $\tilde{r} \equiv \sqrt{x^2+y^2+\gamma z^2}$.  We perform a variable change $x = \cos \theta$:
\begin{align*}
\bullet\bullet &= \sqrt{\gamma}\, \frac{e^{i k_0 z}}{(2\pi)^2v^2}\int\limits_{0}^\infty k^2 dk \int\limits_{-1}^1 dx \frac{e^{i k \tilde{r} x}}{k^2 + \left( \omega_n - i \mu\right)^2/v^2} =
\sqrt{\gamma}\, \frac{e^{i k_0 z}}{2\pi^2v^2}\frac{1}{\tilde{r}}\int\limits_{0}^\infty \frac{k \sin k \tilde{r}}{k^2 + \left( \omega_n - i \mu\right)^2/v^2} dk = \\
&= \sqrt{\gamma}\, \frac{1}{4\pi v^2}\cdot \frac{e^{i k_0 z}}{\sqrt{x^2+y^2+\gamma z^2} } \cdot e^{-\frac{1}{v} \left( \omega_n - i \mu\right) \sgn\omega_n \sqrt{x^2+y^2+\gamma z^2} }
\end{align*}
Similarly, we can treat the second Weyl point $k_z = -k_0$, and eventually we have
\begin{equation}
I(\bs{r}) = I_+(\bs{r}) + I_-(\bs{r}) = \frac{1}{2\pi v^2}\,\sqrt{\gamma}\, \frac{\cos k_0 z}{\sqrt{x^2+y^2+\gamma z^2}} e^{-\frac{1}{v}(\omega_n - i\mu)\sgn\omega_n \sqrt{x^2+y^2+\gamma z^2}} 
\label{AppIrformula}
\end{equation}
Substituting the expressions in Eqs.~(\ref{AppbulkG}), (\ref{AppdefI0xyz}), and (\ref{AppIrformula}) into Eq.~(\ref{BubblediagramBulk}) we get the final expression for the bulk RKKY contribution:
\begin{align}
\nonumber \mathcal{E}_b(\bs{r}) = 4 \Big\{ 2 J_\perp^2 \left( S_{1x}S_{2x} + S_{1y}S_{2y} \right) \left(\Sigma_0 + \Sigma_z + 2 \Sigma_{0z} \right) + \phantom{aaaaaaaaaaaaaaaaaaaaaaaaaaaaaaaaa}\\
 S_{1z}S_{2z} \left[ \left( \Sigma_0 + \Sigma_z - \Sigma_{x^2+y^2} \right) J_z^2 + 4 \Sigma_{0z} J_z \delta J_z + \left( \Sigma_0 + \Sigma_z + \Sigma_{x^2+y^2} \right) \delta J_z^2 \right]  \Big\},
\label{AppRKKYbulk}
\end{align}
with $\Sigma_0, \Sigma_z, \Sigma_{0z}$ and $\Sigma_{x^2+y^2}$ defined in Appendix \ref{AppMatsubaraFrequencySummation}. In the limit of $\mu,\,T \to 0$ we get:
\begin{align}
\nonumber &\mathcal{E}_b(\bs{r})\Big|_{\mu=T=0} = \frac{\gamma}{4\pi^3v^3R^{11}} 
\Bigg\{ \Big[2 J_\perp^2 \left(\bs{S}_1\cdot \bs{S}_2 -S_{1z}S_{2z}\right)\left( f(x,z) -v^2 R^6 \right) + \\
\nonumber &S_{1z}S_{2z} \left[ J_z^2 \left( f(x,z) + v^2R^4\left(4x^2-\gamma z^2 \right) \right) + \delta J_z^2 \left(f(x,z) -v^2R^4\left(6x^2+\gamma z^2 \right) \right) \right] \Big] \cos^2 k_0 z\, + \\
&\Big[2 J_\perp^2 \left(\bs{S}_1\cdot \bs{S}_2 -S_{1z}S_{2z}\right) +  
 (J_z^2+\delta J_z^2)S_{1z}S_{2z} \Big] \cdot 10 m_1^2 R^2 k_0z \sin k_0 z \Big[2R^2 k_0z \sin k_0 z - (2x^2-5\gamma z^2)\cos k_0z \Big]\Bigg\}
\end{align}
where we defined $R \equiv \sqrt{x^2+\gamma z^2}$ and $f(x,z) \equiv m_1^2 \left(5x^4-25\gamma x^2 z^2 +33 \gamma^2z^4\right)$.
 
\section{Summation over Matsubara frequencies}
\label{AppMatsubaraFrequencySummation}
We start by computing two auxiliary sums:
\begin{equation}
\mathcal{S}_1^{0}\equiv T\sum\limits_{\omega_n} e^{-\alpha \left(\omega_n - i \mu \right)\sgn \omega_n} = \frac{T}{\sh \pi \alpha T} \cos \alpha\mu \quad\text{and}\quad \mathcal{S}_2^{0} \equiv T\sum\limits_{\omega_n} \sgn \omega_n e^{-\alpha \left(\omega_n - i \mu \right)\sgn \omega_n} = i \frac{T}{\sh \pi \alpha T} \sin \alpha\mu,
\label{auxsums}
\end{equation}
where $\alpha \equiv \frac{2}{v}\sqrt{x^2+y^2+\gamma z^2}$. Using derivation under summation sign we get the sets of sums:
\begin{align}
&\mathcal{S}_1^{1}\equiv T\sum\limits_{\omega_n} \left(\omega_n - i \mu \right)\sgn \omega_n e^{-\alpha \left(\omega_n - i \mu \right)\sgn \omega_n} = \frac{T}{\sh \pi \alpha T} \left[ \mu \sin \alpha\mu + \pi T \cth \pi \alpha T \cdot \cos\alpha\mu \right] \\
&\mathcal{S}_1^{2}\equiv T\sum\limits_{\omega_n} \left(\omega_n - i \mu \right)^2 e^{-\alpha \left(\omega_n - i \mu \right)\sgn \omega_n} = \frac{T}{\sh \pi \alpha T} \left\{ \left[\pi^2T^2\left(1 + \frac{2}{\sh^2\pi\alpha T} \right) -\mu^2 \right] \cos\alpha\mu + 2\pi \mu T \cth \pi \alpha T \cdot \sin \alpha\mu\right\} \\
\nonumber &\mathcal{S}_1^{3}\equiv T\sum\limits_{\omega_n} \left(\omega_n - i \mu \right)^3 \sgn\omega_n e^{-\alpha \left(\omega_n - i \mu \right)\sgn \omega_n} = \\
&= \frac{T}{\sh \pi \alpha T} \left\{ \pi T \cth \pi \alpha T \cdot \left[\pi^2T^2\left(1 + \frac{6}{\sh^2\pi\alpha T} \right) - 3 \mu^2 \right] \cos\alpha\mu + \mu \left[ 3\pi^2T^2\left(1 + \frac{2}{\sh^2\pi\alpha T} \right) - \mu^2 \right] \sin \alpha\mu\right\} \\
\nonumber &\mathcal{S}_1^{4}\equiv T\sum\limits_{\omega_n} \left(\omega_n - i \mu \right)^4 e^{-\alpha \left(\omega_n - i \mu \right)\sgn \omega_n} = \\
\nonumber &= \frac{T}{\sh \pi \alpha T} \left\{ \left[\pi^4T^4\left(1 + \frac{20}{\sh^2\pi\alpha T} + \frac{24}{\sh^4\pi\alpha T}\right) - 6 \pi^2 \mu^2 T^2 \left(1 + \frac{2}{\sh^2\pi\alpha T} \right) + \mu^4 \right] \cos\alpha\mu + \right. \\
&\left. + 4\pi\mu T \cth \pi \alpha T \cdot \left[ \pi^2T^2\left(1 + \frac{6}{\sh^2\pi\alpha T} \right) - \mu^2 \right] \sin \alpha\mu\right\},
\end{align}
whereas the second sum in Eq.~(\ref{auxsums}) yields:
\begin{align}
&\mathcal{S}_2^{1}\equiv T\sum\limits_{\omega_n} \left(\omega_n - i \mu \right) e^{-\alpha \left(\omega_n - i \mu \right)\sgn \omega_n} = i\frac{T}{\sh \pi \alpha T} \left[ -\mu \cos \alpha\mu + \pi T \cth \pi \alpha T \cdot \sin\alpha\mu \right] \\
&\mathcal{S}_2^{2}\equiv T\sum\limits_{\omega_n} \left(\omega_n - i \mu \right)^2 \sgn \omega_n e^{-\alpha \left(\omega_n - i \mu \right)\sgn \omega_n} = i\frac{T}{\sh \pi \alpha T} \left\{ \left[\pi^2T^2\left(1 + \frac{2}{\sh^2\pi\alpha T} \right) -\mu^2 \right] \sin\alpha\mu - 2\pi \mu T \cth \pi \alpha T \cdot \cos \alpha\mu\right\} \\
\nonumber &\mathcal{S}_2^{3}\equiv T\sum\limits_{\omega_n} \left(\omega_n - i \mu \right)^3 e^{-\alpha \left(\omega_n - i \mu \right)\sgn \omega_n} = \\
&= i\frac{T}{\sh \pi \alpha T} \left\{ \left[\mu^2 - 3\pi^2T^2\left(1 + \frac{2}{\sh^2\pi\alpha T} \right) \right] \mu\cos\alpha\mu + \pi T \cth \pi \alpha T \left[ \pi^2T^2\left(1 + \frac{6}{\sh^2\pi\alpha T} \right) - 3 \mu^2 \right]\cdot \cos \alpha\mu\right\}
\end{align}

Using the auxiliary sums computed above, we can now find different parts of Eq.~(\ref{RKKYbulk}):
\begin{align}
&\Sigma_0 \equiv  T\sum\limits_{\omega_n} I_0^2 = -\frac{\gamma}{(2\pi v^2)^2}\frac{\cos^2 k_0 z}{x^2+y^2+\gamma z^2} \mathcal{S}_1^{2} \\
&\Sigma_{x^2+y^2}  \equiv  T\sum\limits_{\omega_n} \left( I_x^2 + I_y^2 \right) = -\frac{\gamma}{(2\pi v^2)^2} \frac{(x^2+y^2)\cos^2 k_0 z}{(x^2+y^2+\gamma z^2)^2} \left[ \mathcal{S}_1^{2}  + \frac{2v}{\sqrt{x^2+y^2+\gamma z^2}}  \mathcal{S}_1^{1} + \frac{v^2}{x^2+y^2+\gamma z^2} \mathcal{S}_1^{0} \right] \\
&\Sigma_z \equiv T\sum\limits_{\omega_n} I_z^2 = \frac{\gamma m_1^2}{(2\pi v^2)^2} \left[\frac{A^2}{v^4} \mathcal{S}_1^4 + \frac{2AB}{v^3} \mathcal{S}_1^3 +\frac{B}{v^2} \left(B + \frac{2A}{\sqrt{x^2+y^2+\gamma z^2}} \right)\mathcal{S}_1^2 + \frac{2}{v}\frac{B^2}{\sqrt{x^2+y^2+\gamma z^2}} \mathcal{S}_1^1 + \frac{B^2}{x^2+y^2+\gamma z^2} \mathcal{S}_1^0 \right] \\
&\Sigma_{0z} \equiv T\sum\limits_{\omega_n} I_0I_z = -\frac{i \gamma m_1}{(2\pi v^2)^2} \frac{\cos k_0 z}{\sqrt{x^2+y^2+\gamma z^2}} \left( \frac{A}{v^2} \mathcal{S}_2^3 + \frac{B}{v} \mathcal{S}_2^2 + \frac{B}{\sqrt{x^2+y^2+\gamma z^2}} \mathcal{S}_2^1 \right),
\end{align}
where we defined
$$
A \equiv \frac{\gamma z^2}{(x^2+y^2+\gamma z^2)^{3/2}} \cos k_0 z, \quad B \equiv \frac{2k_0 z(x^2+y^2+\gamma z^2)\sin k_0 z - (x^2+y^2-2\gamma z^2)\cos k_0 z}{(x^2+y^2+\gamma z^2)^2}.
$$
Finally, from Eq.~(\ref{RKKYbulk}) we obtain the expression for the bulk RKKY interaction term:
\begin{align}
\nonumber \mathcal{E}_b(\bs{r}) = 4 \Big\{ 2 J_\perp^2 \left( S_{1x}S_{2x} + S_{1y}S_{2y} \right) \left(\Sigma_0 + \Sigma_z + 2 \Sigma_{0z} \right) + \phantom{aaaaaaaaaaaaaaaaaaaaaaaaaaaaaaaaaaaaaa} \\
+ S_{1z}S_{2z} \left[ \left( \Sigma_0 + \Sigma_z - \Sigma_{x^2+y^2} \right) J_z^2 + 4 \Sigma_{0z} J_z \delta J_z +  \left( \Sigma_0 + \Sigma_z + \Sigma_{x^2+y^2} \right) \delta J_z^2 \right]  \Big\}
\end{align}

\section{RKKY interaction: superconducting case}\label{AppRKKYSC}
We proceed by considering a proximity induced superconducting minigap in the surface states of the Dirac semimetal \cite{Zyuzin2014}. 
For example, such a gap might be realized within the normal region of the SNS junction performed on the surface of a given semimetal.
The Bogoliubov-de Gennes Hamiltonian of the surface states in the Nambu representation reads
\begin{equation}
\mathcal{H}_{s,SC}(\mathbf{k}) = \bigg[
 \begin{matrix}
 \mathcal{H}_{s}(\bs{k}) -\mu  & \Delta \\
 \Delta & - \mathcal{H}_s(\bs{k}) +\mu
 \end{matrix}
\bigg].
\end{equation}
Here we consider the gap to be a spatially homogeneous positive constant. Performing calculations similar to those for the normal state we obtain the energy of the RKKY interaction between two localized spins in the zero temperature limit in the form
\begin{eqnarray}\label{RKKY_SC}
\mathcal{E}_{s,SC} \left(\bs{r} \right) = \mathcal{E}_s\left(\bs{r}\right)|_{T=0}\,F_1\left(2|x|/\xi \right) + \frac{J_z^2}{\pi v^2} \Delta S_{1z}S_{2z} \times \frac{4m^2_1 \left( \sin k_0 z - k_0 z \cos k_0 z\right)^2}{\pi^2 v^2 z^6} F_2 \left(2|x|/\xi\right),~~~
\end{eqnarray}
where $\mathcal{E}_s\left(\bs{r} \right)$ is the interaction energy in the normal state given by Eq.~(\ref{RKKYsurface}), and $\xi = v/\Delta$ is the superconducting coherence length of the surface states. The asymptotes of the functions $F_1(p) = \int_{p}^{\infty}dt \sqrt{1-(p/t)^2} e^{-t}$ and $F_2(p) = \int_{p}^{\infty}dt \frac{p e^{-t}}{t  \sqrt{t^2 - p^2} }$  are given by $F_1(p)=1$ and $F_2(p)=\pi/2$ at $p\ll 1$, while  one has $F_1(p) = F_2(p) = \sqrt{\frac{\pi}{2p}} e^{-p}$ at
$p\gg 1$.

As expected, when $\Delta \to 0$ (or equivalently, $\xi \to \infty$) the expression above coincides with Eq.~(\ref{RKKYsurface}). The second term in Eq.~(\ref{RKKY_SC}) does not depend on the position of the chemical potential provided the dispersion of the surface states is linear. We thus find that the superconducting coupling gives rise to antiferromagnetic interactions between localized spins which is expected for s-wave pairing. Therefore, the RKKY energy is minimized if the localized spins lie in the $y=0$ plane and point in opposite directions along the $z$ axis provided $|J_z| \gg |J_{\perp}|$.

\section{Bulk-surface interference terms}
\label{AppBulkSurfaceInterference}

In order to consider bulk-surface interference terms affecting the RKKY exchange energy, we write the full Green's function as a sum of surface and bulk contributions: $G = G_s \otimes \tau_0 + G_b$. We note that we need to add the orbital degree of freedom $\tau_0$ into the surface Green's function to match its dimension with that of the bulk one, and we will omit it below for brevity. The cross-terms can be then written as follows: 
\begin{equation}
\mathcal{E}_{s-b} = T \sum_{i,j;n} \tr \left[ J_i S_{1i}\sigma_i  G_s(\omega_n,\bs{r}) J_j S_{2j} \sigma_j G_b(\omega_n, -\bs{r})\right]  +  T  \sum_{i,j;n} \tr \left[ J_i S_{1i}\sigma_i  G_b(\omega_n,\bs{r})J_j S_{2j} \sigma_j  G_s(\omega_n, -\bs{r})\right]
\end{equation}
The calculation can be performed along the same lines as the calculations above. We present here the final result:
\begin{align}
\nonumber \mathcal{E}_{s-b} = \frac{4\sqrt{\gamma}}{\pi v^3} \frac{\cos k_0 z}{\sqrt{x^2+\gamma z^2}} &\frac{2m_1 \left( \sin k_0 z - k_0 z \cos k_0 z\right)}{\pi v z^3} \Bigg[ \left(\sum\limits_{i=1}^3 J_i^2 S_{1i}S_{2i}\right) \frac{T}{\sh \pi \tilde{\alpha} T} \left(\mu \sin \tilde{\alpha} \mu + \pi T \cth \pi \tilde{\alpha} T \cdot \cos \tilde{\alpha} \mu \right) +  \\
& + \sgn x \,J_x J_y \left(S_{1x}S_{2y}-S_{2x}S_{1y} \right) \frac{T}{\sh \pi \tilde{\alpha} T} \left(-\mu \cos \tilde{\alpha} \mu + \pi T \cth \pi \tilde{\alpha} T \cdot \sin \tilde{\alpha} \mu \right) \Bigg],
\end{align}
where $\tilde{\alpha} \equiv \left(|x|+\sqrt{x^2+z^2} \right)/v$. From the equation above we can derive the sought-for limits of $T \to 0$ at $\mu = 0$ and $\mu \neq 0$. It is easy to derive that at the charge neutrality point $\mathcal{E}_{s-b} \propto 1/r^6$, whereas at $\mu \neq 0$ we have $\mathcal{E}_{s-b} \propto 1/r^5$.

\end{document}